\documentclass[aps,prd,nofootinbib,twocolumn,preprintnumbers,floatfix,superscriptaddress,longbibliography]{revtex4-1}
\usepackage[colorlinks, linkcolor=red, anchorcolor=blue, citecolor=blue, colorlinks=true, urlcolor=blue]{hyperref}
\usepackage{amssymb}
\usepackage{amsmath}
\usepackage{amsthm}
\usepackage{mathrsfs}
\usepackage{threeparttable}
\usepackage{gensymb}
\usepackage{lipsum}
\usepackage[figuresright]{rotating}
\usepackage{fontawesome}
\usepackage{slashed}
\usepackage{graphicx}
\usepackage{indentfirst}
\usepackage{subfigure}
\usepackage{ulem}
\usepackage{bm}
\usepackage{multirow}
\usepackage{epstopdf}
\usepackage{breakurl}
\usepackage{xspace}
\usepackage{cancel}
\makeatother
\allowdisplaybreaks

\begin{document}
    
\title{QCD axion bubbles in the presence of ALP resonant conversion} 
 
\preprint{BNU-23-028} 

\author{Hai-Jun Li} 
\email{lihaijun@itp.ac.cn}
\affiliation{Key Laboratory of Theoretical Physics, Institute of Theoretical Physics, Chinese Academy of Sciences, Beijing 100190, China}
\affiliation{Center for Advanced Quantum Studies, Department of Physics, Beijing Normal University, Beijing 100875, China}

\date{\today}

\begin{abstract}

The QCD axion bubbles can form due to an explicit breaking of the Peccei-Quinn symmetry in the early Universe. 
In this paper, we investigate the modified formation of QCD axion bubble in the presence of axionlike particle (ALP), considering its resonant conversion to QCD axion. 
We consider a general scenario where the QCD axion mixes with ALP before the QCD phase transition. 
In this scenario, the energy density of the ALP can be adiabatic transferred to the QCD axion at a temperature $T_R$, resulting in the suppression of the cosmic background temperature $T_B$ at which the energy density of the QCD axion equals that of the radiation. 
The QCD axion bubbles form when the QCD axions arise during the QCD phase transition. 
Finally, we briefly discuss the impact of the formation of QCD axion bubbles on the formation of primordial black holes.


\end{abstract}
\maketitle

\section{Introduction}

The strong CP problem in the Standard Model (SM) is a long-standing problem and can be solved by the Peccei-Quinn (PQ) mechanism with a spontaneously broken $U(1)$ PQ symmetry \cite{Peccei:1977hh, Peccei:1977ur}.
The PQ mechanism predicted a light pseudo Nambu-Goldstone (NG) boson, axion (also called the QCD axion) \cite{Weinberg:1977ma, Wilczek:1977pj}, which acquires a tiny mass from the QCD non-perturbative effects \cite{tHooft:1976rip, tHooft:1976snw}.
When the potential of the QCD axion is generated by the QCD instanton, the axion is stable at the minimum value of CP conservation, which solves the strong CP problem.
The QCD axion is the potential cold dark matter (DM) candidate if non-thermally produced in the early Universe through the misalignment mechanism \cite{Preskill:1982cy, Abbott:1982af, Dine:1982ah, Preskill:1982cy}.
The QCD axion is massless at high temperatures, as the cosmic temperature decreases, it acquires a non-zero mass during the QCD phase transition and starts to oscillate when its mass becomes comparable to the Hubble parameter, which explains the observed DM abundance.
See $\rm e.g.$ Refs.~\cite{DiLuzio:2020wdo, Chadha-Day:2021szb, Adams:2022pbo} for recent reviews.

The PQ symmetry is supposedly broken before or during inflation.
In this case, the QCD axion is massless during inflation and acquires quantum fluctuations.
To suppress the isocurvature perturbations, a feasible method is considering an explicit PQ symmetry breaking \cite{Dine:2004cq, Jeong:2013xta, Takahashi:2015waa, Harigaya:2015soa}.
The PQ symmetry is an approximate global symmetry and can be strongly broken in the early Universe with the multiple vacua \cite{Kallosh:1995hi, Banks:2010zn, Witten:2017hdv, Lee:2018yak, Harlow:2018jwu, Yin:2020dfn, Jeong:2022kdr}.
The QCD axion bubbles \cite{Kitajima:2020kig} can form due to this explicit PQ symmetry breaking.
In this case, axion acquires a light mass and oscillates when this mass is greater than the Hubble parameter.
Note that the explicit PQ symmetry breaking potential for the axion will disappear before the QCD phase transition.
Therefore, the final QCD axion abundance can also be calculated through the misalignment mechanism.
However, the initial misalignment angle is determined by the explicit PQ symmetry breaking, which can be split into different values.
If the axion initial value is smaller than a critical value, the axion is stabilized near the origin.
On the contrary, the axion will be stabilized at another minimum if the axion initial value is larger than the critical value.
During the QCD phase transition, the QCD axions start to oscillate near the origin and another large value.
The former case can account for the cold DM abundance, while the latter one can form the high axion density region, which is called the QCD axion bubble.
The concept of QCD axion bubbles was first proposed in Ref.~\cite{Kitajima:2020kig}, which is similar to the baryon bubbles in the inhomogeneous Affleck-Dine baryogenesis \cite{Dolgov:2008wu, Hasegawa:2018yuy}.
Additionally, they also investigated the formations of the primordial black holes (PBHs) and axion miniclusters from the QCD axion bubbles.

In this paper, we investigate the modified QCD axion bubble formation in the presence of the axionlike particle (ALP) to QCD axion resonant conversion.
We introduce a general case that the QCD axion mixes with ALP before the QCD phase transition.
In this case, the ALP to QCD axion resonant conversion is considered to take place at the temperature $T_R$, and the ALP energy density can be adiabatic transferred to the QCD axion. 
We find that this will lead to the suppression of the cosmic background temperature $T_B$ in which the QCD axion energy density is equal to the radiation energy density.
The QCD axion bubbles are formed when the QCD axions arise during the QCD phase transition.
Finally, we briefly discuss this impact on PBHs formation in the QCD axion bubbles scenario, leading to the enhancement of the minimum PBH mass.

The rest of this paper is organized as follows.
In Sec.~\ref{sec_QCD_axion}, we briefly review the QCD axion DM and the misalignment mechanism. 
In Sec.~\ref{sec_axion_bubbles}, we investigate the effect on QCD axion bubble formation in the presence of the ALP resonant conversion, and also the impact on PBHs formation in the QCD axion bubbles scenario.
Finally, the conclusion is given in Sec.~\ref{sec_conclusion}.

\section{QCD axion and misalignment mechanism}
\label{sec_QCD_axion} 
 
Here we briefly review the QCD axion DM and the misalignment mechanism.
The QCD axion is a pseudo NG boson with a spontaneously broken $U(1)$ PQ symmetry.
It couples to gluons with the following effective Lagrangian
\begin{eqnarray}
\mathcal{L}_{agg}=-\frac{\alpha_s}{8\pi}\frac{\phi}{f_a}G^{a\, \mu\nu}\tilde{G}_{\mu\nu}^a\, ,
\end{eqnarray}
where $\alpha_s$ is the strong fine structure constant, $\phi$ is the QCD axion field, $f_a=v_a/N_{\rm DW}$ is the axion decay constant, $v_a$ is the spontaneous breaking scale of the PQ symmetry, $N_{\rm DW}$ is the number of domain wall, $G^{a\, \mu\nu}$ and $\tilde{G}_{\mu\nu}^a$ are the gluon field strength tensor and dual tensor, respectively.
The resulting effective potential of the QCD axion is given by
\begin{eqnarray}
V_{\rm QCD}(\phi)=m_a^2(T) f_a^2\left[1-\cos\left(\frac{\phi}{f_a}\right)\right]\, ,
\label{eq_Va}
\end{eqnarray}
where $m_a(T)$ is the temperature-dependent QCD axion mass for $T\gtrsim T_{\rm QCD}$ ($\sim150\, \rm MeV$) \cite{Borsanyi:2016ksw}
\begin{eqnarray}
m_a(T)\simeq m_{a,0}\left(\frac{T}{T_{\rm QCD}}\right)^{-4.08}\, ,
\end{eqnarray}
with the zero-temperature axion mass \cite{GrillidiCortona:2015jxo}
\begin{eqnarray}
m_{a,0}\simeq 5.70(7)\,{\mu \rm eV}\left(\frac{f_a}{10^{12}\,{\rm GeV}}\right)^{-1}\, .
\end{eqnarray}  

In the misalignment mechanism \cite{Preskill:1982cy, Abbott:1982af, Dine:1982ah, Preskill:1982cy}, as the cosmic temperature decreases the QCD axion starts to oscillate when its mass $m_a(T)$ becomes comparable to the Hubble parameter $H(T)$
\begin{eqnarray}
3H(T_a)=m_a(T_a)\, ,
\end{eqnarray}
then we have the oscillation temperature 
\begin{eqnarray}
T_a\simeq0.96\, {\rm GeV}\left(\frac{g_*(T_a)}{61.75}\right)^{-0.082}\left(\frac{f_a}{10^{12}\, \rm GeV}\right)^{-0.16}\, ,
\label{eq_QCD_T_oscillation}
\end{eqnarray}
where $g_*(T)$ is the number of effective degrees of freedom of the energy density.
The QCD axion number density at $T_a$ is given by
\begin{eqnarray}
n_a(T_a)=\frac{1}{2}m_a(T_a)f_a^2\left\langle\theta_i^2f(\theta_i)\right\rangle\chi \, ,
\end{eqnarray}
where $\theta_i$ is the initial misalignment angle, $\chi\simeq1.44$ is a numerical factor \cite{Turner:1985si}, and $f(\theta_i)$ is the anharmonic factor
\begin{eqnarray}
f(\theta_i)\simeq\left[\ln\left(\frac{e}{1-\theta_i^2/\pi^2}\right)\right]^{1.16}\, ,
\end{eqnarray}
which is taken as $f(\theta_i)\simeq1$ for $|\theta_i|\ll\pi$ \cite{Lyth:1991ub, Visinelli:2009zm}.
The present axion energy density $\rho_a(T_0)= m_{a,0} n_a(T_0)$ is
\begin{eqnarray}
\rho_a(T_0)=\frac{m_{a,0} m_a(T_a) s(T_0)}{2s(T_a)} f_a^2\left\langle\theta_i^2f(\theta_i)\right\rangle\chi \, ,
\end{eqnarray}
where $s(T)=2\pi^2 g_{*s}(T)T^3/45$ is the entropy density, $g_{*s}(T)$ is the number of effective degrees of freedom of the entropy density, and $T_0$ is the present CMB temperature.
Then we have the current QCD axion abundance $\Omega_ah^2=\rho_a(T_0)/\rho_c h^2$ as
\begin{eqnarray}
\begin{aligned}
\Omega_ah^2&\simeq0.14 \left(\frac{g_{*s}(T_0)}{3.94}\right)\left(\frac{g_*(T_a)}{61.75}\right)^{-0.42}\\
&\times\left(\frac{f_a}{10^{12}\, \rm GeV}\right)^{1.16}\left\langle\theta_i^2f(\theta_i)\right\rangle\, ,
\label{eq_misa_Omega} 
\end{aligned}
\end{eqnarray}
where $\rho_c=3H_0^2M_{\rm Pl}^2$ is the critical energy density, $M_{\rm Pl}\simeq2.44\times10^{18}\, \rm GeV$ is the reduced Planck mass, and $h\simeq0.68$ is the reduced Hubble constant.
In order to explain the observed cold DM abundance, $\Omega_{\rm DM}h^2\simeq0.12$ \cite{Planck:2018vyg}, we derive the initial misalignment angle
\begin{eqnarray}
\begin{aligned}
\theta_i&\simeq0.87\left(\frac{g_{*s}(T_0)}{3.94}\right)^{-1/2}\left(\frac{g_*(T_a)}{61.75}\right)^{0.21}\\
&\times\left(\frac{f_a}{10^{12}\, \rm GeV}\right)^{-0.58}\, ,
\end{aligned}
\end{eqnarray}
which is valid for $f_a\lesssim10^{17}\, \rm GeV$.
 
\section{QCD axion bubbles in the presence of ALP resonant conversion}
\label{sec_axion_bubbles} 

In this section, we investigate the effect on QCD axion bubble formation in the presence of the ALP to QCD axion resonant conversion, and also the impact on PBHs formation from the QCD axion bubbles.

\subsection{QCD axion bubbles}

The QCD axion bubbles can form due to an explicit PQ symmetry breaking in the early Universe, and are formed during the QCD phase transition.
There are many scenarios for this explicit PQ symmetry breaking, such as the Witten effect of monopoles in hidden sectors \cite{Nomura:2015xil, Kawasaki:2015lpf, Kawasaki:2017xwt}, a larger scale of the spontaneous PQ symmetry breaking with the higher dimensional term \cite{Chiba:2003vp, Takahashi:2003db, Higaki:2014ooa, Co:2019jts} and the hidden non-Abelian gauge interactions \cite{Takahashi:2015waa}, and a stronger QCD with the large Higgs field expectation value \cite{Choi:1996fs, Jeong:2013xta}.
In Ref.~\cite{Kitajima:2020kig}, they considered the Witten effect as an example of the explicit PQ symmetry breaking to form the QCD axion bubbles.
Considering a large axion decay constant $f_a\sim\mathcal{O}(10^{16}-10^{17})\, \rm GeV$ and the axion is stabilized at the potential minima with
\begin{eqnarray}
\phi_{\rm min}^0\simeq0\, , \quad \phi_{\rm min}^1\simeq \pi f_a\, , \quad \cdot\cdot\cdot \, , 
\end{eqnarray}
which corresponds to the effective initial misalignment angle $\theta_{i,n}$ with
\begin{eqnarray}
\theta_{i,0}=0-\theta_i\, , \quad \theta_{i,1}= \pi -\theta_i \, , \quad \cdot\cdot\cdot \, .
\end{eqnarray} 
When the QCD axion $V_{\rm QCD}(\phi)$ arises during the QCD phase transition, the state $\phi_{\rm min}^0$ with the effective initial angle $\theta_{i,0}$ explains the cold DM abundance, corresponding to a small initial misalignment angle
\begin{eqnarray}
\begin{aligned}
\theta_i&\simeq4.29\times10^{-3}\left(\frac{g_{*s}(T_0)}{3.94}\right)^{-1/2}\left(\frac{g_*(T_a)}{61.75}\right)^{0.21}\\
&\times\left(\frac{f_a}{10^{16}\, \rm GeV}\right)^{-0.58}\, .
\label{eq_theta_i_2} 
\end{aligned}
\end{eqnarray}
On the other hand, if the initial value of axion is larger than a critical value $\phi_{\rm crit}$, the axion will settle down into the minimum $\phi_{\rm min}^1$ with the initial angle $\theta_{i,1}$.
In this case, since the large initial misalignment angle $\pi$, the local axion density at the minimum $\phi_{\rm min}^1$ becomes much higher than that at $\phi_{\rm min}^0$, which forms the high density QCD axion bubbles.
When the axion dominates the radiation in the bubbles, $\rm i.e.$, the axion energy density is equal to the radiation energy density, we can define the cosmic background temperature, $T_B$.
The local axion energy density in the bubbles at $T_B$ is given by
\begin{eqnarray}
\rho_{a,B}=\frac{m_{a,0} m_a(T_a) s(T_B)}{2s(T_a)} f_a^2\left\langle\theta_{i,1}^2f(\theta_{i,1})\right\rangle\chi \, ,
\end{eqnarray}
and the radiation energy density is defined as
\begin{eqnarray}
\rho_{R,B}=\frac{\pi ^2}{30} g_*(T_B)T_B^4 \, .
\end{eqnarray}
Considering $\rho_{a,B}=\rho_{R,B}$ and the effective initial angle $\theta_{i,1}= \pi-\theta_i$, we have the temperature
\begin{eqnarray}
\begin{aligned}
T_B&\simeq2.13\times10^{-2}\, {\rm MeV}\times\left(\frac{g_*(T_a)}{61.75}\right)^{-0.42}\\
&\times\left(\frac{f_a}{10^{16}\, \rm GeV}\right)^{1.16}\left\langle\theta_{i,1}^2f(\theta_{i,1})\right\rangle\chi \, .
\label{eq_TB}
\end{aligned}
\end{eqnarray}
Substituting Eq.~(\ref{eq_theta_i_2}) into Eq.~(\ref{eq_TB}), we can derive $T_B\simeq 2.84 \, {\rm MeV}$ for $f_a\sim\mathcal{O}(10^{16})\, \rm GeV$.
              
\begin{figure}[t]
\centering
\includegraphics[width=0.47\textwidth]{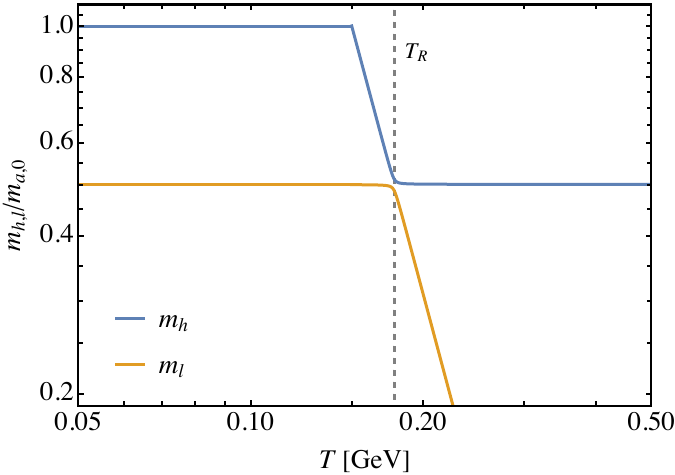}
\caption{Illustration of the ALP to QCD axion resonant conversion at the temperature $T_R$.
The solid lines distinctly represent the normalized temperature-dependent mass eigenvalues, where $m_h(T)$ and $m_l(T)$ denote the heavy and light mass eigenvalues, respectively, as they evolve with temperature. 
Notice that the cosmic temperature decreases from right to left in the plot.
This graphical representation provides insight into the dynamic interplay between the ALP and QCD axion masses during the resonant conversion.
Here we set $f_a=10^{16}\, \rm GeV$, $f_a/f_A=20$, and $m_{a,0}/m_A=2$.}
\label{fig_eigenvalue}
\end{figure} 

\subsection{ALP to QCD axion resonant conversion} 
 
\begin{figure*}[t]
\centering
\includegraphics[width=0.49\textwidth]{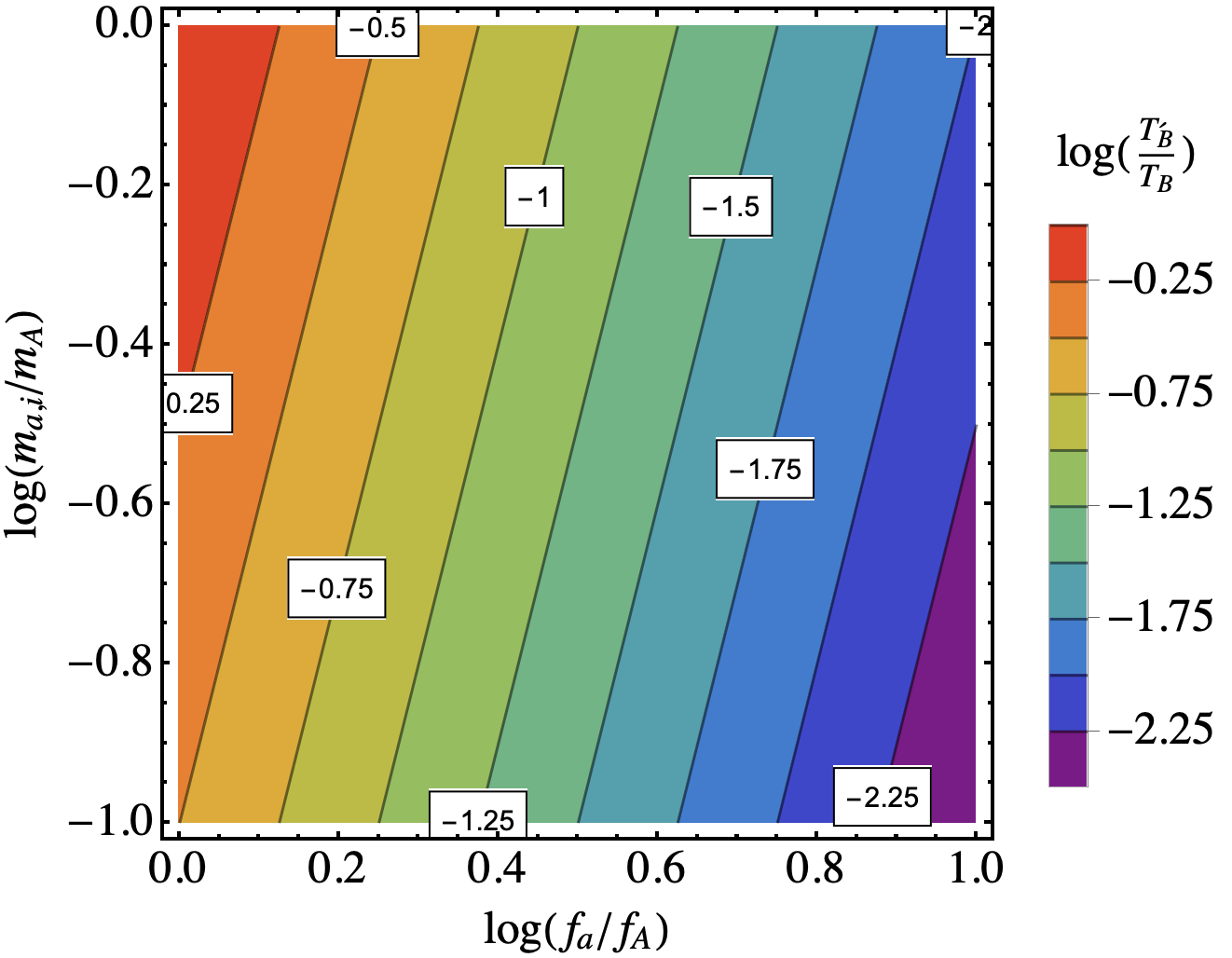}\quad\includegraphics[width=0.49\textwidth]{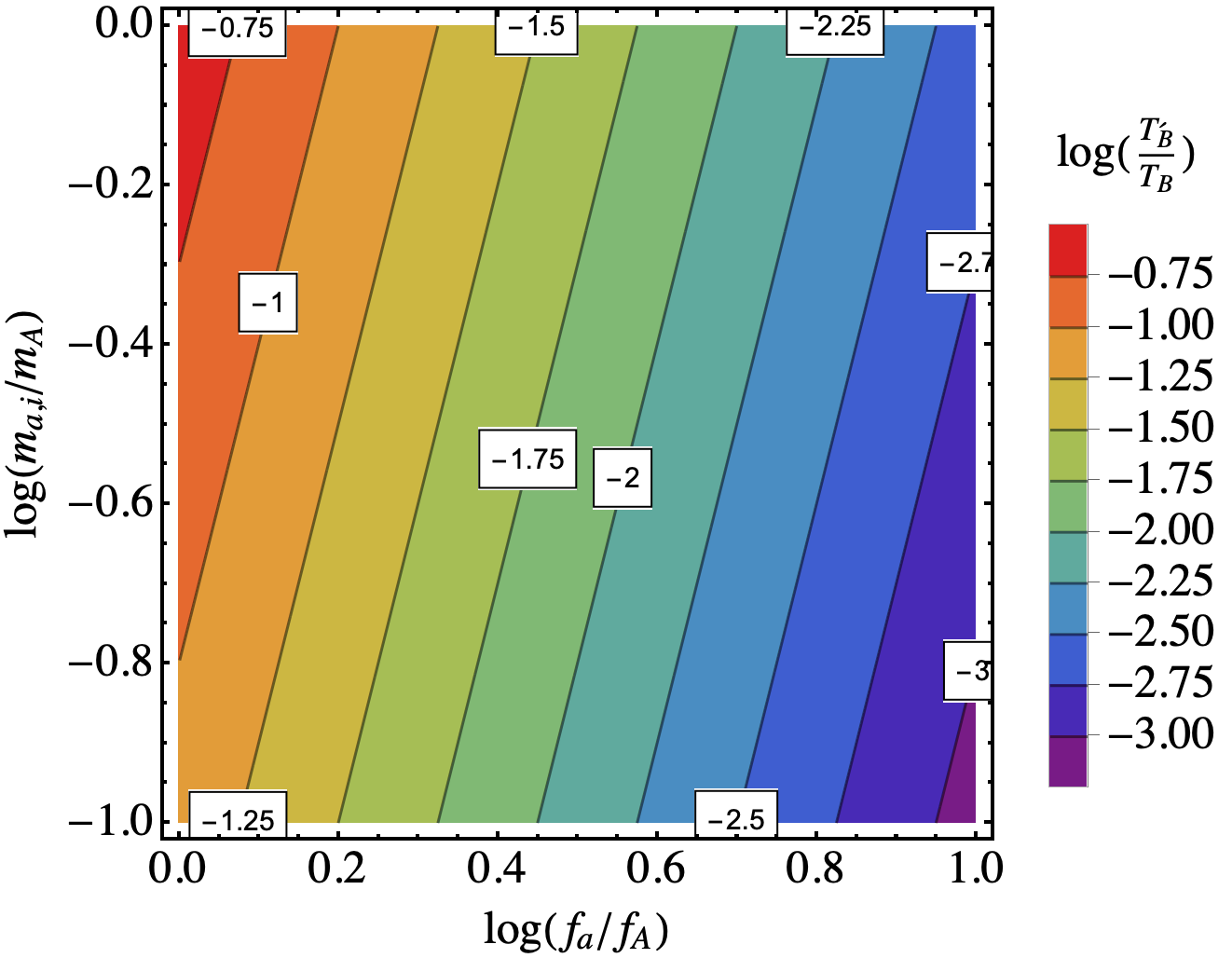}
\caption{Variations in the temperature ratio $T'_B/T_B$, defined by Eq.~(\ref{ratio_TB}), across the $\{\log(f_a/f_A), \log(m_{a,i}/m_A)\}$ parameter space.
Left panel: displays the distribution when a specific set of parameter $\theta_{i,A} = \pi$ is applied.
Right panel: shows the distribution for an alternative set of parameter $\theta_{i,A} = \pi/2$.
These panels not only illustrate how the temperature ratio distribution varies within the given parameter range but also highlight the potential for significant suppression of the temperature ratio under different parameter conditions.}
\label{fig_TB}
\end{figure*}
 
In the following, we investigate the effect in the presence of the ALP to QCD axion resonant conversion.  
We consider a general case that the QCD axion mixes with ALP with the mixing potential \cite{Ho:2018qur, Li:2023xkn} 
\begin{eqnarray}
\begin{aligned}
V_{\rm mix}(\phi, \varphi)&=m_A^2 f_A^2\left[1-\cos\left(\frac{\phi}{f_a}+\frac{\varphi}{f_A}+\delta\right)\right]\\
&+m_a^2(T) f_a^2\left[1-\cos\left(\frac{\phi}{f_a}\right)\right]\, ,
\label{eq_Vmix}
\end{aligned}
\end{eqnarray}    
where $\varphi$ is the ALP field, $m_A$ and $f_A$ are the ALP mass and the decay constant, respectively, and $\delta$ is the CP phase.   
For simplicity, in this context we assume a temperature-independent ALP mass\footnote{Given that ALP does not have a specific model detailing its mass variation with temperature, unlike the QCD axion, we have chosen, for the sake of simplicity and in accordance with prevalent practices in the literature, to adopt a constant-temperature mass model for ALP in our analysis.} and take the CP phase $\delta$ to be zero\footnote{This can be regarded as equivalent to requiring an independent solution to the strong CP problem. Since this assignment has no bearing on the evolution of the two axion fields during their mixing, it consequently leaves the axion energy density in our scenario unaltered.}.
The mass mixing matrix is given by
\begin{eqnarray}
\mathbf{M}^2=
\left(
\begin{array}{cc}
m_a^2(T)+\dfrac{m_A^2 f_A^2}{f_a^2}  & \quad \dfrac{m_A^2 f_A}{f_a} \\
\dfrac{m_A^2 f_A}{f_a} & \quad m_A^2
\end{array}
\right)\, ,
\end{eqnarray}
then we can derive the heavy and light mass eigenvalues $m_{h, l}(T)$.
See Fig.~\ref{fig_eigenvalue} for the illustration of the normalized mass eigenvalues as functions of the cosmic temperature $T$.          
Here we consider the conditions under which the ALP to QCD axion resonant conversion can take place: 
\begin{eqnarray}
\dfrac{f_a}{f_A}> 1\, , \quad \dfrac{m_{a,0}}{m_A}> 1\, .
\end{eqnarray}
In this case, the resonant conversion occurs at the temperature $T_R$, which is given by  
\begin{eqnarray}
m_{a}(T_R)\simeq m_A\, .
\end{eqnarray}
At the temperature $T>T_R$, the light mass eigenstate $m_l$ is associated with the QCD axion, whereas below $T_R$, it is the heay mass eigenstate $m_h$ that comprises the QCD axion. 
Conversely, at the temperature above $T_R$, the heay mass eigenstate $m_h$ pertains to the ALP, whereas at the temperature below $T_R$, it is the light mass eigenstate $m_l$ that forms the ALP.
The axion energy transition at $T_R$ is considered to be adiabatic, which can be roughly satisfied when
\begin{eqnarray}
T_a \gg T_R\, ,
\end{eqnarray}
where $T_a$ is the QCD axion oscillation temperature.

To obtain the energy density of the QCD axion at the temperature $T'_B$, we should begin with the ALP field at high temperatures.
Its initial energy density at the oscillation temperature $T_{i,A}$ is given by
\begin{eqnarray}
\rho_{A,i}=\dfrac{1}{2}m_A^2 f_A^2 \theta_{i, A}^2\, ,
\end{eqnarray}
where $\theta_{i, A}$ is the initial misalignment angle of the ALP. 
At $T_R<T<T_{i,A}$, the ALP energy density is adiabatic invariant.
Using $N_A \equiv \rho_A a^3 /m_A$, where $a$ is the scale factor, we have the ALP energy density at the temperature $T_R$ as
\begin{eqnarray}
\rho_{A,R}=\frac{1}{2}m_A^2 f_A^2 \theta_{i,A}^2 \left(\frac{a_{i,A}}{a_R}\right)^3 \, ,
\end{eqnarray}
where $a_{i,A}$ and $a_R$ correspond to the scale factors at $T_{i,A}$ and $T_R$, respectively.    
At $T_R$, the ALP energy density $\rho_{A,R}$ is adiabatic transferred to the QCD axion $\rho_{a,R}$.  
Then at  $T'_B<T<T_R$, the adiabatic approximation is valid again with $N_a \equiv \rho_a a^3 /m_a$, and we have the QCD axion energy density at $T'_B$ as
\begin{eqnarray}
\rho'_{a,B}=\frac{1}{2}m_{a,0}m_A f_A^2 \theta_{i,A}^2 \left(\frac{a_{i,A}}{a'_B}\right)^3 \, ,
\end{eqnarray} 
where $a'_B$ is the scale factor at $T'_B$.
Note that $T'_B$ is given by $\rho'_{a,B}=\rho'_{R,B}$.
Now compared with the no ALP resonant conversion case   
\begin{eqnarray}
\rho_{a,B}=\frac{1}{2}m_{a,0}m_{a,i} f_a^2 \theta_{i,1}^2 \left(\frac{a_{i,a}}{a_B}\right)^3 \, ,
\end{eqnarray} 
we find that the temperature $T_B$ can be suppressed as
\begin{eqnarray}
\begin{aligned}
\dfrac{T'_B}{T_B}&\simeq\sqrt[4]{\dfrac{m_A f_A^2 \theta_{i,A}^2}{m_{a,i} f_a^2 \theta_{i,1}^2} \left(\frac{a_{i,A} a_B}{a_{i,a} a'_B}\right)^3}\\
&=\dfrac{m_{a,i}^{1/2} f_A^2 \theta_{i,A}^2}{\pi^2 m_A^{1/2} f_a^2}\, ,
\label{ratio_TB}
\end{aligned}
\end{eqnarray}
where $m_{a,i}\equiv m_a(T_a)$.
Note that $a(T) \propto \sqrt{t} \propto 1/T \propto 1/\sqrt{H}$, $a_i(T) \propto 1/\sqrt{H} \propto 1/\sqrt{m_i}$, and $m_i$ corresponds to the axion mass at the oscillation temperature that given by $3H=m_i$. 
In Fig.~\ref{fig_TB}, we show the distributions of this factor across the $\{\log(f_a/f_A), \log(m_{a,i}/m_A)\}$ plane with $\theta_{i,A}=\pi$ and $\pi/2$, respectively.
We find that within the defined range of the parameter space we are examining, the temperature $T_B$ undergoes a notable suppression.
One can further calculate the present QCD axion energy density in this manner to explain the abundance of cold DM, but this is not the focus of the current context and therefore is not presented here. For further details, please refer to Refs.~\cite{Ho:2018qur, Li:2023xkn}.
     
\begin{figure*}[t]
\centering
\includegraphics[width=0.49\textwidth]{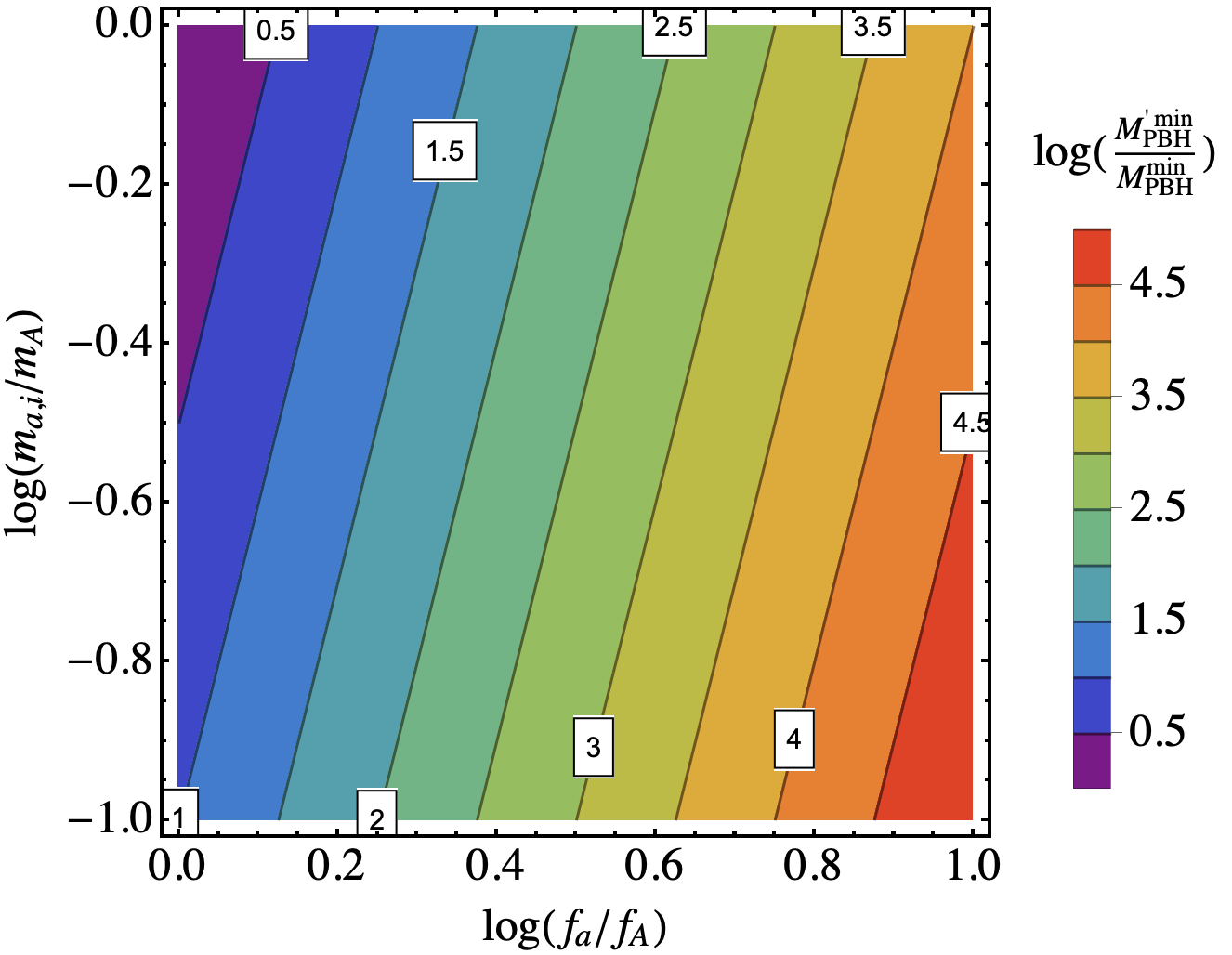}\quad\includegraphics[width=0.49\textwidth]{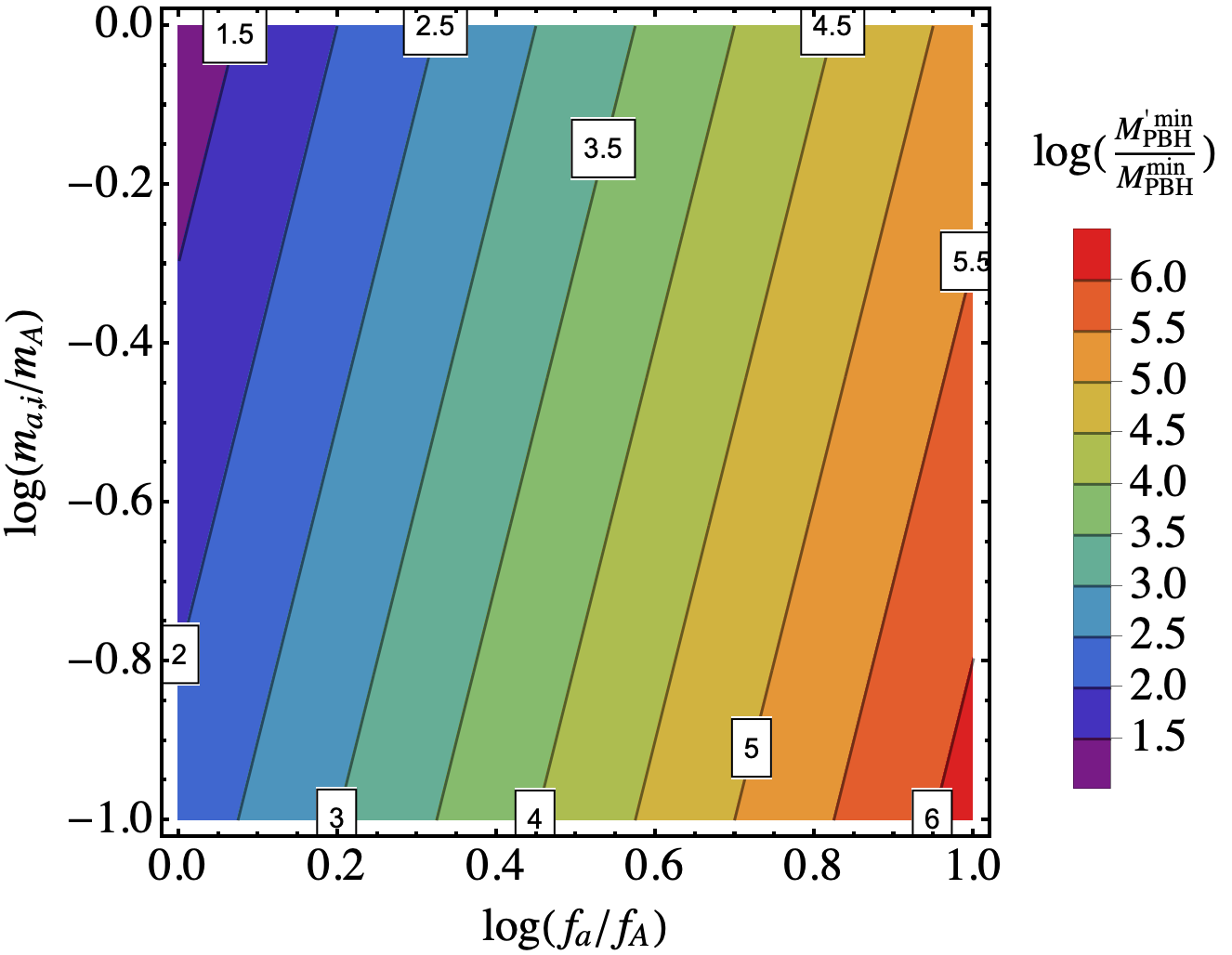}
\caption{Variations in the minimum PBH mass ratio $M_{\rm PBH}^{'\rm min}/M_{\rm PBH}^{\rm min}$, defined by Eq.~(\ref{ratio_Mmin}), across the $\{\log(f_a/f_A), \log(m_{a,i}/m_A)\}$ parameter space.
Left panel: we set $\theta_{i,A}=\pi$.
Right panel: we set $\theta_{i,A}=\pi/2$.
These panels not only demonstrate the variability in the minimum PBH mass ratio but also emphasize the potential for substantial mass enhancement under specific parameter configurations.}
\label{fig_Mmin}
\end{figure*}      
     
\subsection{Primordial black holes}
\label{sec_PBH}

In this subsection, we provide a concise overview of the formation of PBHs from the QCD axion bubbles. 
PBHs, which can arise from significant density perturbations in the early Universe, are also compelling candidates for DM \cite{Bird:2016dcv, Carr:2020xqk, Green:2020jor, Carr:2021bzv, Li:2022mcf}.
At the time of their formation $t_f$, the initial mass of a PBH is described by the formula $M_{\rm PBH}=4\pi\gamma\rho_R/(3H_f^3)$ \cite{Carr:2009jm}, where $\gamma\simeq0.2$ is the gravitational collapse factor \cite{Carr:1975qj}, $\rho_R$ represents the radiation energy density, and $H_f$ is the Hubble parameter at $t_f$.
Subsequently, we derive the PBH mass at the formation time as follows:
\begin{eqnarray}
\begin{aligned}
\frac{M_{\rm PBH}}{M_\odot}&\simeq0.03\times\left(\frac{\gamma}{0.2}\right)\left( \frac{g_*(T_f)}{10.75}\right)^{-1/2}\\
&\times\left( \frac{T_f}{1\, \rm GeV}\right)^{-2}\, ,
\label{eq_pbh}
\end{aligned} 
\end{eqnarray}
where $T_f$ corresponds to the temperature at $t_f$, and $M_\odot$ denotes the solar mass. 
This equation allows us to estimate the PBH mass based on the relevant physical parameters.

In the context of the QCD axion bubbles scenario \cite{Kitajima:2020kig}, PBHs form when axions dominate the radiation inside the bubbles, $\rm i.e.$, when the temperature $T_f$ is less than or equal to the cosmic background temperature $T_B$, and the bubble size exceeds the horizon size\footnote{Note that the fluctuation of ALPs may also have an impact on the production of PBHs in the early Universe. However, since our focus here is on PBHs generated by modified QCD axion bubbles, we have not considered this effect separately. Further discussion on this effect may be included in future work.}. 
When these bubbles enter the horizon, the local energy density within them significantly surpasses the background radiation density.
Given that the mass of a PBH resulting from bubble collapse cannot exceed the background horizon mass by much \cite{Kopp:2010sh, Carr:2014pga}, it is assumed that the PBH mass in the axion bubbles scenario is equivalent to the horizon mass of the background radiation, even when $T_f < T_B$. 
By substituting Eq.~(\ref{eq_TB}) into Eq.~(\ref{eq_pbh}), we derive the minimum PBH mass
\begin{eqnarray}
\begin{aligned}
\frac{M_{\rm PBH}^{\rm min}}{M_\odot}&\simeq6.58\times10^7\left(\frac{\gamma}{0.2}\right)\left(\frac{g_*(T_f)}{10.75}\right)^{-1/2}\\
&\times\left( \frac{g_*(T_a)}{61.75}\right)^{0.84}\left(\frac{f_a}{10^{16}\, \rm GeV}\right)^{-2.33}\\
&\times\left(\left\langle\theta_{i,1}^2f(\theta_{i,1})\right\rangle\chi\right)^{-2}\, .
\label{eq_Mmin}
\end{aligned} 
\end{eqnarray}
For $f_a\sim\mathcal{O}(10^{16})\, \rm GeV$, the minimum PBH mass is approximately $M_{\rm PBH}^{\rm min}\simeq3.71\times10^3{M_\odot}$.
When considering the impact of resonant conversion from ALP to QCD axion, we find that the minimum PBH mass can be enhanced by a factor
\begin{eqnarray}
\frac{M_{\rm PBH}^{'\rm min}}{M_{\rm PBH}^{\rm min}}\simeq \dfrac{\pi^4 m_A f_a^4}{m_{a,i} f_A^4 \theta_{i,A}^4}\, .
\label{ratio_Mmin}
\end{eqnarray}
Fig.~\ref{fig_Mmin} illustrates the distributions of this enhancement factor across the $\{\log(f_a/f_A), \log(m_{a,i}/m_A)\}$ plane, with $\theta_{i,A}$ fixed at $\pi$ and $\pi/2$. 
Notably, these mass-enhanced PBHs could potentially serve as the seeds for supermassive black holes (SMBHs) at high redshifts \cite{Li:2023zyc}.

Additionally, another intriguing phenomenon associated with the formation of QCD axion bubbles is the creation of axion miniclusters \cite{Kitajima:2020kig}. 
These miniclusters are gravitationally bound aggregations of axion DM \cite{Hogan:1988mp, Kolb:1993zz, Fairbairn:2017sil, Xiao:2021nkb, Ellis:2022grh, Dandoy:2022prp}. 
The mass and size of these miniclusters are contingent upon the Hubble volume at the time when the QCD axion begins to oscillate.
It is noteworthy that axion miniclusters can form when the bubbles enter the horizon prior to the axions becoming the dominant component of radiation within the bubbles. 
This condition sets the stage for the gravitational binding of axions, leading to the emergence of these miniclusters.

\section{Conclusion}
\label{sec_conclusion}
 
In summary, our investigation has focused on the modified formation of QCD axion bubbles in the context of resonant conversion between ALP and QCD axion. 
The formation of these bubbles can be attributed to an explicit breaking of the PQ symmetry in the early Universe. 
We have introduced a generalized scenario where the QCD axion can mix with ALP prior to the QCD phase transition. 
Within this framework, resonant conversion from ALP to QCD axion is considered to occur at a temperature $T_R$, with the ALP energy density being adiabatic transferred to the QCD axion under specific conditions.
Our findings reveal that this resonant conversion process leads to a suppression of the cosmic background temperature $T_B$ at which the QCD axion energy density equals the radiation energy density. 
Subsequently, during the QCD phase transition, the emergence of QCD axions triggers the formation of axion bubbles.
Furthermore, we have discussed the implications of this scenario for the formation of PBHs. 
Our analysis suggests that the minimum PBH mass can be increased to a certain extent, and these PBHs may potentially serve as the seeds for SMBHs. 


\medskip\noindent{\bf Acknowledgments.}---%
The author would like to thank Wei Chao, Naoya Kitajima, Shota Nakagawa, Fuminobu Takahashi, and Yu-Feng Zhou for helpful discussions and valuable comments.
This work was partly supported by the National Natural Science Foundation of China (NSFC) (Grants No.~11775025 and No.~12175027), and partly supported by the Key Laboratory of Theoretical Physics in Institute of Theoretical Physics, CAS.

 

\bibliography{references}
\end{document}